# ATLAS Detector Status and Physics Startup Plans

M. Bosman  (on behalf of the ATLAS collaboration)
*IFAE, Barcelona, E-08193, Spain*

The status of the ATLAS detector shortly before the start-up of the LHC is presented.  The progress in the commissioning of the detector, as well as plans for early physics analysis are outlined.

## 1.  INTRODUCTION

ATLAS (A Toroidal LHC ApparatuS) is one of the two general purpose detectors built to probe proton-proton collisions at 14 TeV. Here we present the status and physics startup plans of ATLAS, as of beginning of August 2008. This paper is organized as follows. The ATLAS detector concept is explained, followed by a brief history of the installation. The current status of the detector and the commissioning work preparing for first collisions are described. Then the plans for early physics are outlined.

## 2.  THE ATLAS DETECTOR

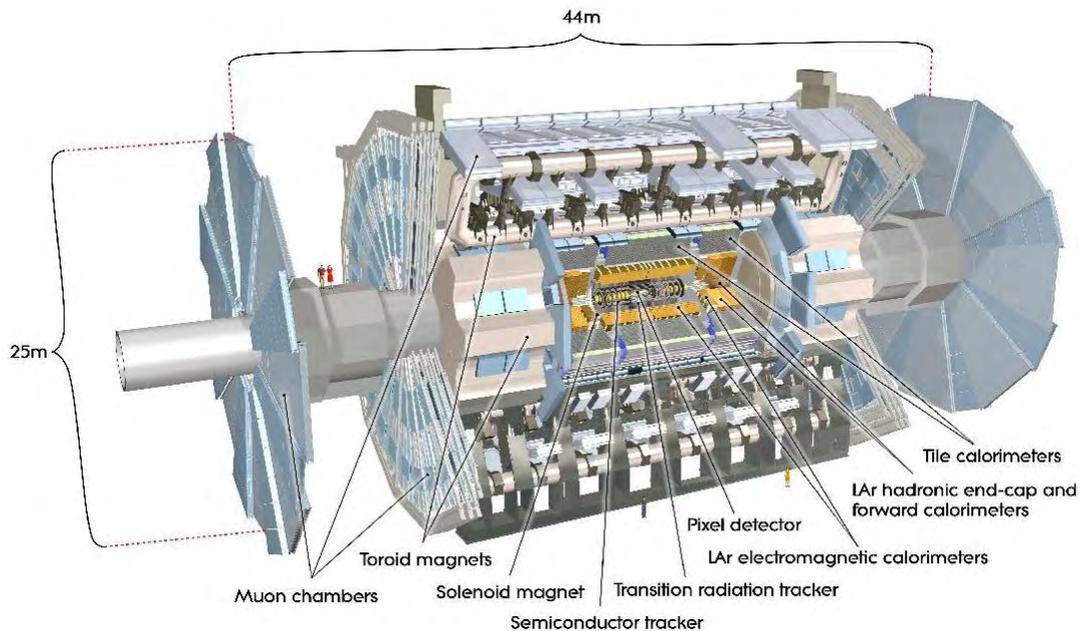

Figure 1: Layout of the ATLAS detector.

A schematic layout of ATLAS [1] is shown in Figure 1. The ATLAS detector presents the typical large acceptance concentric collider detector structure.  The inner tracking detectors (ID) surround the beam pipe. They consist of cylindrical layers (in the barrel part) and disks (in the forward parts) of silicon pixel and strips, followed by a Transition Radiation Detector (TRT) used for tracking and particle identification ($e/\pi$ separation). These detectors are immersed in a 2 Tesla magnetic field generated by the solenoid coil placed in front of the electromagnetic lead-liquid argon sampling calorimeter system with accordion shape (LAr) housed in a barrel and two endcap cryostats.  The endcap cryostats contain also a copper absorber hadronic sampling calorimeter as well as the forward calorimeters with electromagnetic and hadronic sections made out of copper and tungsten absorbers respectively. The liquid argon calorimeter is complemented by the barrel and the two extended barrels iron-scintillating tiles sampling hadronic calorimeters (Tilecal). The calorimetry is surrounded in-turn by the large air-core toroid muon system with eight giant 25 m long superconducting barrel coils and the two endcap toroid cryostats containing eight coils each. This system is equipped with precision drift tube chambers (MDT) and multiwire proportional chambers with cathodes segmented into strip (CSC) in the forward region. Resistive Plate Chambers (RPC) and thin-gap (TGC) provide trigger capability in the barrel and in the endcap region respectively. The overall size and shape of the detector are those of a 45 m long cylinder of 24 m diameter situated in the cavern about 100 m underground. In addition, a set of forward detectors contribute to the luminosity measurement. LUCID, a Cherenkov detector, is situated at 17 m from the intersection point close to the beam pipe, while two additional calorimeters will be placed at 140 and 240 m respectively.

## 3. BRIEF HISTORY OF DETECTOR INSTALLATION

The cavern was made available empty in June 2003. The detector elements were then lowered inside the cavern via 60 m long shafts (the widest shaft with a 18 m diameter) and then assembled. The barrel toroid was the first to be completed with the last coil put in place in August 2005. In parallel, the LAr cryostats were lowered in the cavern and the Tilecal modules were mounted around to form the barrels. By May 2006, the central and both extended barrels were completed and had been moved to their location inside the toroid (see Figure 2, left). At the same time, the barrel toroid structure and the big endcap wheels were progressively equipped with muon chambers (see Figure 2, right). The small muon endcap wheels were mounted on surface and lowered through the shafts, as well as the toroid endcap cryostats. The process culminated with the installation of the last muon chamber in July 2008. The inner detector was installed in various phases: TRT and SCT barrel in August 2006, endcaps in May 2007 and Pixel in June 2007 (see Figure 3).

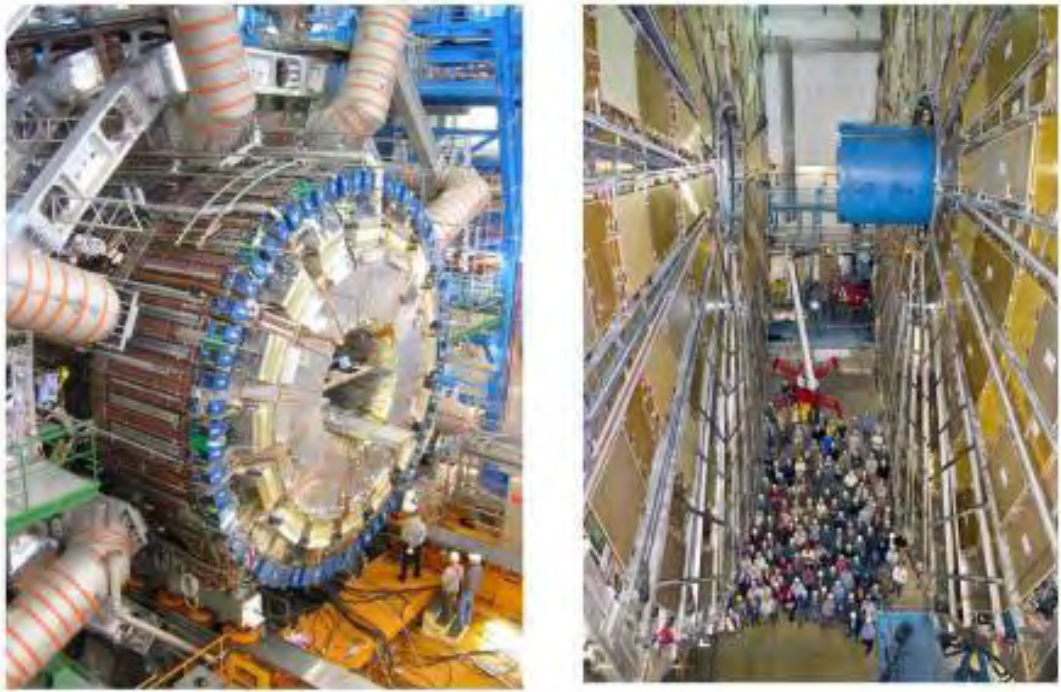

Figure 2: Left: the barrel calorimeter being moved into position inside the toroid in November 2005; Right: the two large endcap muon wheels after completion of the installation in September 2007.

## 4. HARDWARE COMMISSIONING AND STATUS

After installation, the detectors are connected to their readout electronics, operated with their respective cooling, gas or other specific hardware and integrated in the global ATLAS Trigger and Data Acquisition System (TDAQ) [2]. The full chain is then commissioned. In the case of the LAr calorimeters the process was completed in May 2008, after refurbishment of some of the electronics [3]. Since then, the complete calorimeter is up and in steady running mode with the 190,000 channels read out. Only few isolated channels (~ 0.02%) are dead. The control of the power supply of a half-barrel module (~1.5% dead channels) was lost[1]. Similarly, the commissioning of the 10,000 Photomultipliers reading the 5000 cells of Tilecal was completed in May 2008 after refurbishment of part of the electronics. There are ~0.2% of isolated dead cells. One extended barrel module (0.2%) cannot be powered and will be repaired during the first shutdown.

The commissioning of the inner detector was initiated on the surface and continued once installed in the cavern [4]. The solenoid field mapping was done with a precision in $\Delta B/B \sim 10^{-4}$. The TRT is operational and integrated in TDAQ since last year. Delivery of all readout elements is currently being completed. The amount of dead channels is ~ 2%. The detector will be operated with Xenon, for efficient Transition Radiation detection capability, once the gas system will be fully commissioned. Only few weeks of running with the cooling system were possible for the sign-off tests of the SCT. The amount of dead channels is ~ 0.3%, except for one endcap wheel with an additional ~ 1.3% due to a cooling loop problem. The short 6-day period scheduled for Pixel sign-off in April 2008 was interrupted by an incident of the cooling plant situated outside the cavern. The inner detector volume had to be sealed to keep up with the overall ATLAS installation schedule (see Figure 3, right). The cooling plant repair was completed in July 23rd just on-time for the beam pipe bake-out that was done successfully during the last days of July. The amount of dead channels in the Pixel system is ~ 0.4% except for one endcap wheel with 4% dead channels and potentially an additional 8% if the cooling loop problem cannot be solved. The Pixel system will resume its commissioning and run in stand-alone for 4 weeks before joining common ATLAS running, while the SCT will join for limited periods.

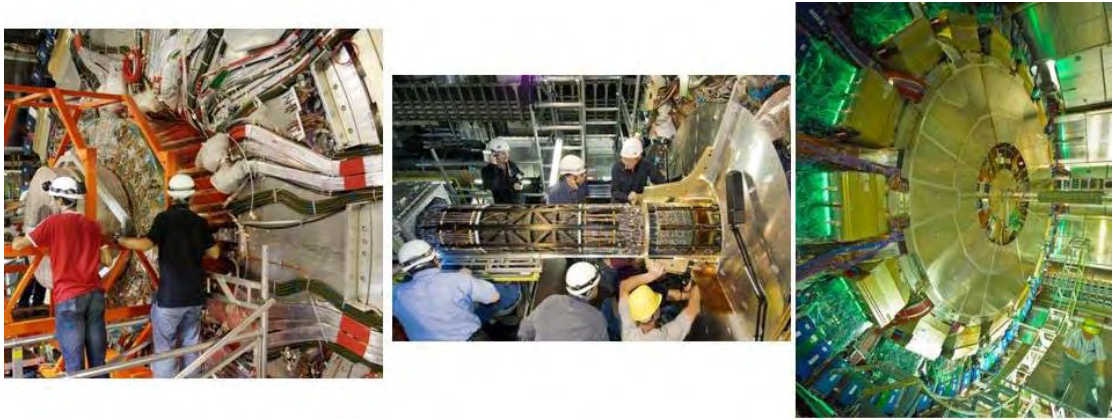

Figure 3: Left: the barrel TRT and SCT being installed in August 2006; Middle: the Pixel being inserted in June 2007; Right: the ID volume sealed with the complex end-plate with 1000 feed-throughs in April 2008.

All muon chambers are installed except for a few chambers staged until 2009 [5]. All wheels are in their final position. Most of the optical alignment system is operational and a precision of ~ 200 μm was already achieved. The magnetic field in the muon spectrometer is monitored using a grid of ~ 1700 Hall probes precisely mounted on the MDT chambers and continuously measuring the field. As of this writing, 98% of these probes are operational. Based on probe readings and magnetic simulations, the field integral is expected to be known to ~ 1.5% at startup, and ~ 0.5% after full commissioning. There are a few chambers with problems resulting from an overpressure accident or from gas leaks, while only very few individual channels are bad. Overall there is no acceptance hole in the muon system, only some loss of redundancy. The connections of the barrel RPC are being finalized, as well as the alignment of a few chambers. All chambers are operated already with their nominal gas.

---

[1] N.B. At the time of this writing, the control of the power supply had been recovered.

The commissioning of the toroids and solenoid magnet system was done in two phases. The solenoid was brought up to full field at the nominal current value of 7.73 kA in August 2006, while the barrel toroid reached the nominal 20.5 kA in November 2006. The procedure was interrupted during the commissioning of the calorimeters and was resumed in June 2008 with a successful combined test of the two barrel magnets. One endcap toroid was then successfully tested, while a leak in an electrical pipe isolator developed during the test of the second endcap on May 23rd. The toroid had to be warmed, repaired and by July 20th it was cooled again and reached 21 kA. The important test of powering simultaneously the three magnets, in view of the strong mechanical forces at play, was done on July 31st up to 15 kA[2].

The Trigger and Data Acquisition system is being exercised since about two years [6]. The Trigger comprises three levels. The Level-1 is hardware based and uses coarse granularity data from the calorimeter and the muon system. Its maximum capacity is currently 75 kHz. The Level-2 and Event Filter are software based and they further reduce the rate of selected events to ~ 200 Hz corresponding to a volume of 300 Mbyte/s sent to the Computer Center for storage. The installation of the nodes in the various PC farms for buffering and building events and for the two high level triggers is now being completed. They are connected to the high bandwidth data network and the control network. The online DAQ software and the slow control system are operational. The system has been tested to nominal rates in dedicated technical runs. The TDAQ system is being used successfully for data taking in detector commissioning runs.

## 5. TOWARDS DATA TAKING: COSMIC MUONS

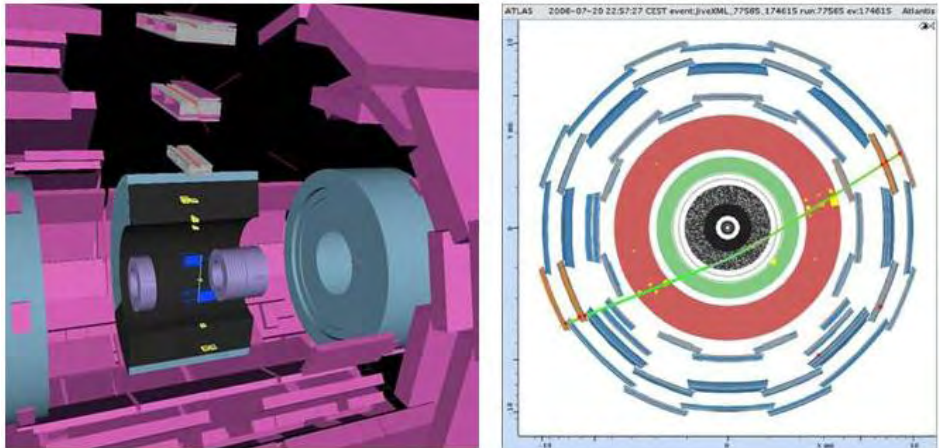

Figure 4: Left: a cosmic ray crossing the detector, the detector elements that are hit are drawn; Right: a cosmic crossing the barrel muon system and the TRT.

As part of the commissioning effort, large samples of cosmic rays events have been recorded with the Level-1 muon and calorimeter trigger over a series of dedicated combined commissioning runs. The objective of these week-long "Milestones" runs (Mx) has been to operate the experiment as complete as possible, to maintain stable running, exercise the data flow, run control and configuration of the TDAQ system, the monitoring, the use of the control room as if in data taking mode and the prompt reconstruction at the CERN Tier-0 computer farm. The number and fraction of detectors involved in the test has been increasing with time. The first week (M1) took place in December 2006 and the most recent (M8) in July 2008 [7]. The rate of cosmic events crossing the inner detector is ~ 15 Hz. Figure 4 shows some cosmic rays crossing the detector.

---

[2] N.B. Full current was reached a few days after the conference.

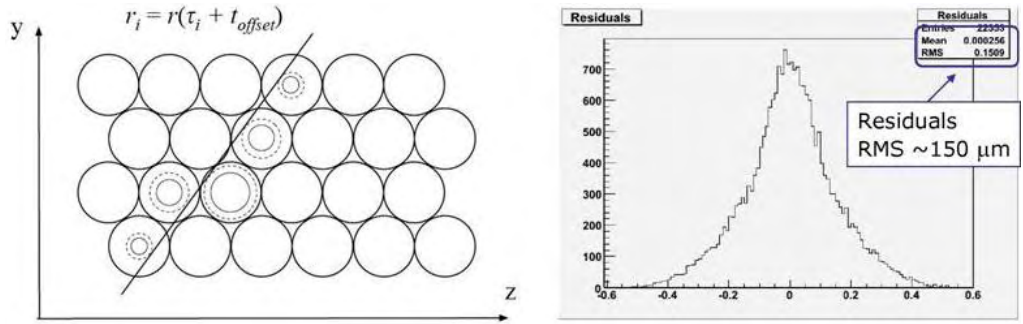

Figure 5: Left: a cosmic ray crossing the MDT tubes; Right: the residual distribution with a RMS of 150 μm.

A wealth of information can be extracted from these events. Here we give a few examples. Figure 5 shows on the left a cosmic crossing a MDT chamber. The time offset and the (r,t) relation can be extracted from the measured residuals. After calibration, the precision on the residuals is ~ 150 μm ( see Figure 5, right), a good step toward the goal of 50 μm precision on the muon sagitta measurement needed to reach 10% accuracy in the momentum measurement of 1 TeV muons. First constraints on the relative alignment between the different detectors can be obtained by comparing the track segments (see Figure 6, left). The precision in the alignment of the MDT and the TRT-SCT achieved so far with cosmics is of ~10 mrad in the two projections.

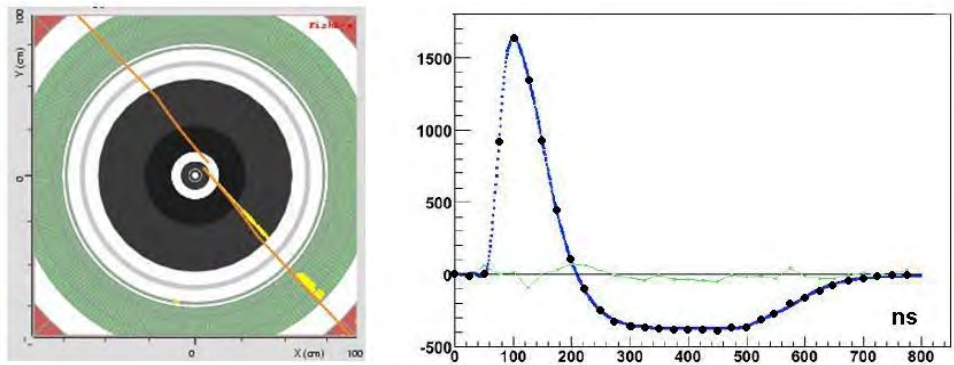

Figure 6: Left: signals deposited by a cosmic muon in the LAr and the ID; Right: signal sampled every 25 ns in the middle layer of the LAr calorimeter (dots) with the predicted pulse shape (line) and residuals.

In the calorimeter, the muons deposit energy according to a Landau distribution. By analysing these distributions, information about detector calibration, uniformity of the response can be extracted. For example, the uniformity of the response of the LAr electromagnetic calorimeter could be cross-checked at the 2% level [3]. The energy measured in Tilecal trigger towers by the Level-1 trigger and in the calorimeter were cross-calibrated. In the LAr calorimeter, large signals due to bremsstrahlung photons emitted by the muons are sometimes observed as seen on the left of Figure 6. The pulse shape of these signals is measured with high precision (see Figure 6, right). These events allowed timing studies (channel intercalibration to ~ 2 ns) and an improved description of the pulse shape that needs to be well understood to reach the ultimate precision. The Tilecal PMTs have been timed so far with a precision of ~ 1ns.

## 6. COLLISIONS: A PHYSICS ROADMAP

With first collisions one of the main goals will be to calibrate the detectors in-situ using well known physics samples. The measurement of known Standard Model signals will serve also to validate and tune MC generators. At the same time, one should be prepared for surprises of very striking new physics signatures. Figure 7 shows a possible physics roadmap as a function of the integrated luminosity. A few examples of analysis are described in this section. A recent and exhaustive review of the physics potential of ATLAS can be found in reference [8].

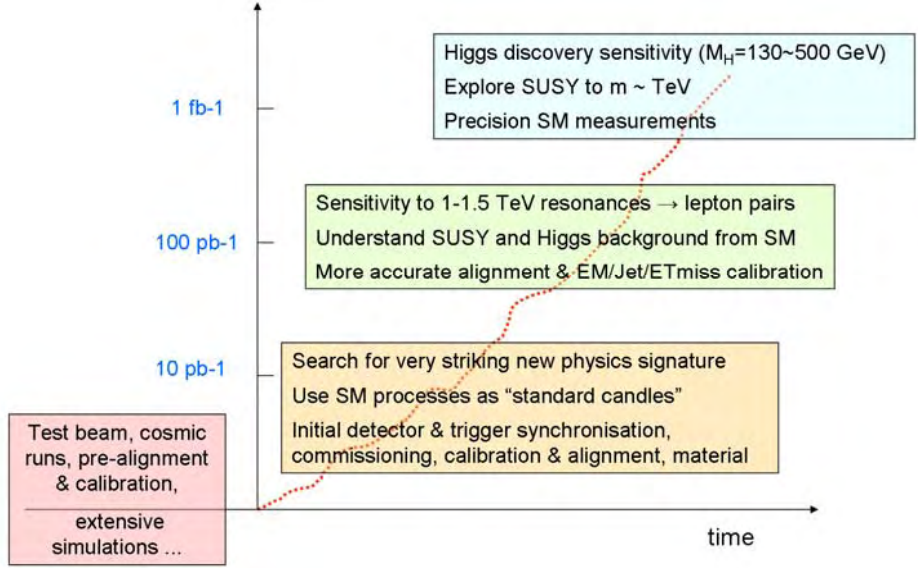

Figure 7: Outline of a physics roadmap as a function of the integrated luminosity.

### 6.1. First Commissioning Signals

At 14 TeV assuming a 30% overall detector and machine efficiency and taking into account all analysis cuts, about 15,000 J/$\Psi$, 2400 $Y$ and 500 Z Bosons decaying to two muons are selected per pb$^{-1}$ (3 days at an initial luminosity of $10^{31}$ cm$^{-2}$ s$^{-1}$) [8]. Similar numbers are expected for electrons. At 10 TeV, cross-sections are typically 30% smaller. From these data, detector efficiencies, the tracker momentum scale, the energy scale of the calorimeters, the lepton trigger and reconstruction efficiencies can be assessed with progressively increasing precision.

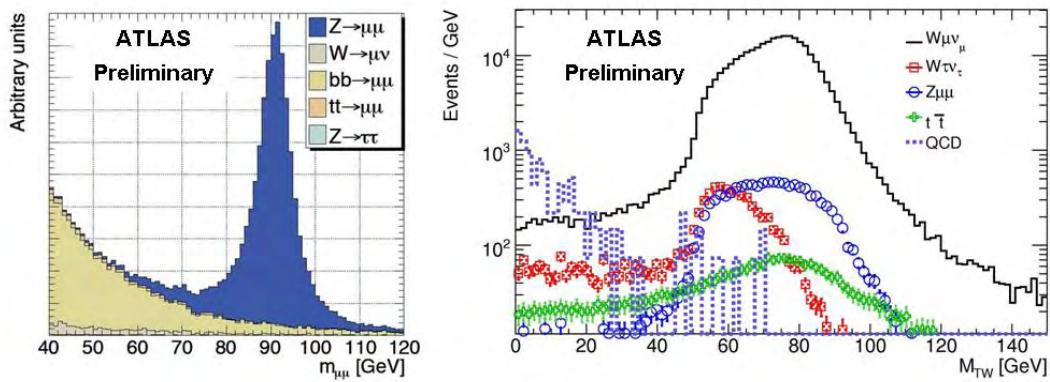

Figure 8: Left: dimuon invariant mass distribution in the Z→$\mu^+\mu^-$ channel, for signal and background, for 50 pb$^{-1}$; Right: transverse mass distribution of the W→$\mu\nu$, for signal and background, for 50 pb$^{-1}$.

## 6.2. W/Z Boson Production

The Z Boson cross-section can be measured with an initial robust analysis, not too demanding on the detector performance, subtracting background from side bands and extracting lepton trigger and reconstruction efficiencies from tag-and-probe methods [8]. The analysis Z→$\mu^+\mu^-$ channel uses the 10 GeV single muon trigger and requests two muon spectrometer tracks with $p_T >$ 20 GeV in the pseudo-rapidity range $|\eta| <$ 2.5, with an invariant mass within 20 GeV of the Z mass. The isolation of the muon is checked in the tracking detector. For an integrated luminosity of 50 pb$^{-1}$, the analysis resulted in about ~ 25,000 signal events and ~ 100 events background (see Figure 8, left). The expected errors on the cross-section are 0.8% (statistical) and 3% (systematic). An additional uncertainty will come from the luminosity, ~ 10% at that stage. To obtain a clean signal of W production in its muon decay mode, one selects a muon track reconstructed in the spectrometer and the inner detector, satisfying $|\eta| <$ 2.5 and $p_T >$ 25 GeV and the 20 GeV single muon trigger. The muon isolation in checked in the calorimeter. A cut of 25 GeV is applied on Missing $E_T$, as well as of 40 GeV on the transverse mass. It is effective against QCD background as shown on the right of Figure 8. About 300,000 events are selected with 20,000 background, resulting in expected errors on the cross-section of 0.2% (statistical) and 3% (systematic). The precision on the measurement of W and Z will be quickly dominated by systematic effects.

## 6.3. Minimum-Bias and QCD

There are large uncertainties in the extrapolation of the cross-section for Minimum-Bias events production from Tevatron to LHC energies, as shown in Figure 9 [8]. It is important to measure that cross-section, as Minimum-Bias events are the source of pile-up at higher luminosity and contribute to the Underlying Event in the hard interaction. They will need to be well understood to do precision physics. The best time to do these measurements is with early data at low instantaneous luminosity. The $p_T$ spectrum of the particles is soft with the cross-section peaking at ~ 250 MeV. To increase the precision of the measurement, the reconstruction of tracks for these events can be extended down to 150 MeV.

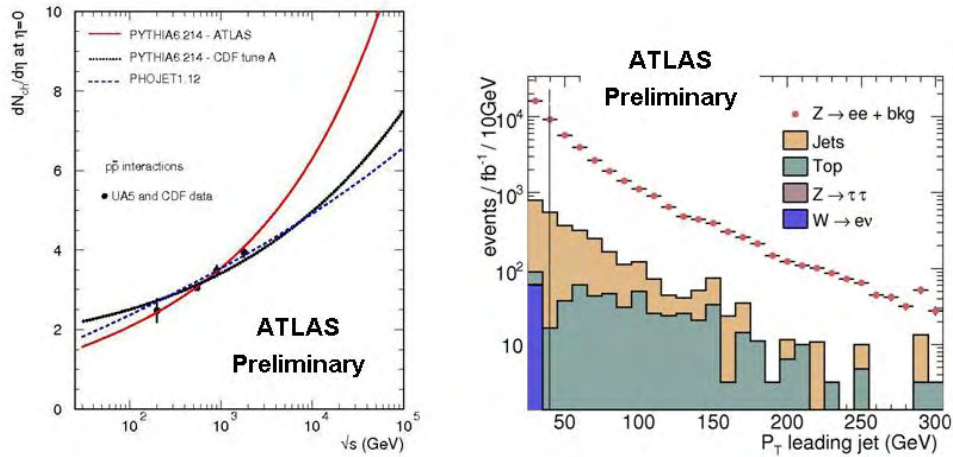

Figure 9: Left: central charged particle density for non-single-diffractive inelastic proton-antiproton collisions as predicted by two different generators Pythia and Phojet tuned to reproduce the Tevatron data; Right: Leading jet $p_T$ distribution in Z→$e^+e^-$+jet events.

When measuring inclusive QCD jet cross-section, one quickly enters in new territory. With an integrated luminosity of 20 pb$^{-1}$ at 14 TeV, ~ 10 jets of 2 TeV will be produced, an energy beyond the reach of the Tevatron. The achievable precision of the measurement will depend on the level of understanding of the Jet Energy Scale at that time. The $\gamma$+jet(s) and Z+jet(s) are useful in-situ jet energy calibration processes, as the $p_T$ balance between the $\gamma$/Z and the jet(s) can be exploited. Figure 9 shows on the right the leading jet $p_T$ distribution observed in Z+jet(s) events. In this analysis, Z Bosons are identified via their decay into two isolated electrons with $p_T > 25$ GeV. In addition, measuring the W/Z+jet(s) cross-section is important, for its intrinsic interest for QCD, and because it represents a relevant background in many new physics searches.

## 6.4. Top

At the LHC, the dominant $t\bar{t}$ pair production process is gluon-gluon fusion which together with the higher center-of-mass energy results in a cross-section about 100 times larger than at the Tevatron, whereas backgrounds are expected to rise only by about a factor 10. The cross-section for the gold-plated semi-leptonic decay $t\bar{t} \rightarrow bWbW \rightarrow b\ell\nu bjj$, with $\ell = e, \mu$, is of the order of 250 pb. A robust analysis has been developed to establish a clear signal with the first data [8]. It relies solely on the measurement of four jets (3 jets with $p_T > 40$ GeV and 1 with $p_T > 20$ GeV), one isolated electron or muon with $p_T > 20$ GeV (including trigger) and Missing $E_T > 20$ GeV. It does not make use of the full b-tagging capability, since precise alignment of the inner detector may not be available in the early days. Figure 10 shows on the left the reconstructed three-jet mass for an integrated luminosity of 100 pb$^{-1}$. A peak with ~ 500 reconstructed top hadronic decays is seen above a background composed mainly of internal combinatorics of $t\bar{t}$ events and W+jets events. Thanks to the over-constrained kinematics of the $t\bar{t}$ system, with the selected events it will be possible to measure the b-tagging performance, the Missing $E_T$, as well as to calibrate the light jet energy scale. The $t\bar{t}$ cross-section will also be measured in the di-lepton channel. With 100 pb$^{-1}$ the cross-section can be measured with an uncertainty of 5-10%, dominated by systematics excluding the uncertainty on luminosity. The production of $t\bar{t}$ could be observed with about ~ 10 pb$^{-1}$.

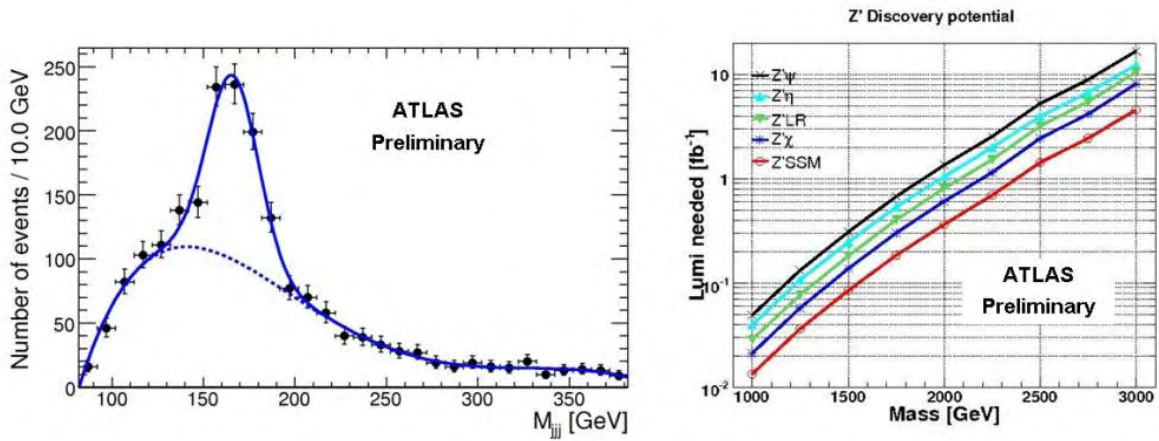

Figure 10: Left: three-jet invariant mass of the selected top events fitted with a polynomial for the background and a Gaussian for the signal; Right: Integrated luminosity needed for 5σ discovery of Z'→ e$^+$e$^−$ as a function of the Z' mass for several theoretical models.

## 6.5. Early Discovery: a Narrow Resonance

New heavy states forming a narrow resonance decaying into opposite sign dileptons are predicted in many extensions of the Standard Model: grand unified theories, technicolor, little Higgs models, and models including extra dimensions [8]. Because of the simplicity of the final state and the clear signature, a mass peak in the invariant mass distribution of opposite sign dileptons on top of the Drell-Yan continuum, these channels are important to study with early data. Ultimate detector performance is not really required. Figure 10 shows on the right the discovery potential of a heavy Z'→ e$^+$e$^−$ as a function of the integrated luminosity. It can be noted that resonance masses above 1 TeV, which are unreachable by the Tevatron experiments, could be discovered with as little as 100 pb$^{-1}$.

## 6.6. Supersymmetry

If Supersymmetry (SUSY) has to do with stabilizing the Higgs mass, new particles are expected at the TeV scale. Some of them, e.g. squarks and gluinos, are strongly interacting and could be produced in pairs with large cross-sections. These particles decay in cascades giving rise to final states with several high-$p_T$ jets, leptons and, in dark-matter motivated scenarios, also large Missing $E_T$ from the escaping lightest stable neutral particles. The signal can observed for example as an excess of events in the tail of the distribution of the effective mass, defined as the scalar sum of the $p_T$ of jets and Missing $E_T$ and eventually leptons (see Figure 10, left), or in the Missing $E_T$ distribution. Potentially the sensitivity is large and gluinos of 1 TeV could be discovered already with 100 pb$^{-1}$. However, the signal can only be firmly established if the detector performance and the backgrounds from QCD, W/Z Boson and top events are well understood. This likely requires more luminosity. Various methods to predict the background distributions have been developed either data-driven, based on MonteCarlo carefully validated with data, or mixed approaches. They should all give consistent results. With more integrated luminosity, one may look for more exclusive signatures, like invariant mass of dileptons. If they are produced in two successive decays in the chain, they will exhibit a characteristic edge with an end point corresponding to the mass difference between the two supersymmetric particles in the decay chain.

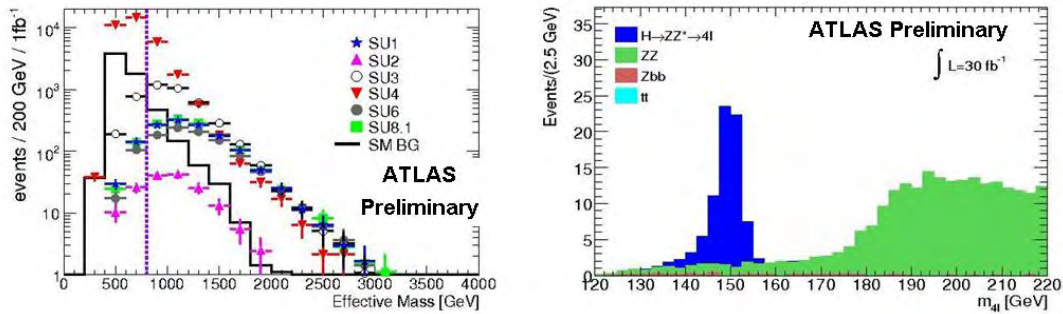

Figure 10: Left: the effective mass distribution for events with four energetic jets, significant Missing $E_T$ and no isolated leptons is shown for a number of SUSY benchmark points and all the backgrounds for 1 fb$^{-1}$ (the final selection cut applied on the effective mass is 800 GeV); Right: Four lepton reconstructed mass distribution for the Higgs signal and background, for a Higgs mass of 150 GeV.

## 6.7. Standard Model Higgs: A More Difficult Case at the Early Stage

In the Standard Model, the electroweak symmetry is spontaneously broken via the Higgs mechanism. A doublet of complex scalar fields is introduced, out of which a single neutral scalar particle, the Higgs boson, remains after symmetry breaking. It is the only piece of the Standard Model that has not been observed experimentally and its possible discovery is one of the main goals of the LHC. The Higgs boson couples preferentially to the heavy particles and this determines its production mechanisms and decay modes. At LHC, gluon fusion, $gg \rightarrow$ H, dominates, followed by vector boson fusion, $qq \rightarrow qq$ H, with two forward quark jets and lack of color exchange between those quarks. Other mechanisms are associated production with weak gauge bosons, $q\bar{q} \rightarrow$ WH/ZH, or heavy quarks, $gg, qq \rightarrow t\bar{t}$ H. They are relevant for low mass Higgs searches. The most favourable decay modes are WW and ZZ when kinematically allowed. At low mass the $b\bar{b}$ decay mode dominates but is difficult to exploit because of the large background from jets (this final state is considered in conjunction with the $t\bar{t}$ H associated production). The most powerful channel with the cleanest signature is H $\rightarrow$ ZZ* $\rightarrow 4\ell$ with good discovery potential in the range 130 < $m_H$ < 600 GeV, except in the mass region around 2$m_W$ (see Figure 10, right). The low mass range will be the most difficult to explore. Below 140 GeV the H $\rightarrow \gamma\gamma$ decay mode is one of the interesting channel that can be exploited both in inclusive analysis or optimized for specific associated production mechanisms. Sensitivity to low mass Higgs comes also from H → ττ in the vector boson fusion mode. Eventually the full mass range will be explored at the LHC. A Standard Model Higgs boson with mass above 130 GeV can be discovered at the 5 sigma level with 10 fb$^{-1}$. More may be needed if the mass is close to the direct LEP bound of 114.4 GeV.

## 7. CONCLUSION

The LHC operation is about to start. During the last five years, all the elements of the ATLAS detector have been progressively installed in the cavern, the last ones in July 2008. Commissioning of the complete subsystems and their integration in the overall Trigger and Data Acquisition System is currently under way. Overall the detector is ready and well functioning with full solid angle coverage and only a very small fraction of dead channels. Many dedicated combined commissioning runs recording cosmic ray events have taken place in the last two years, helping in pre-calibrating the detector and having the full chain ready for data taking. With first collisions the most urgent task will be to understand the detector in details and perform first measurements of Standard Model physics: minimum-bias events, QCD jets, W/Z and top pair production. These measurements will test the Standard Model in a much extended kinematic region and provide important first constraints on the MC generators. One should be open to the possibility of finding new physics if Nature has chosen a scenario which would provide spectacular signatures at LHC energies. We will progressively study the TeV scale in more details with increased statistics looking for hints of the Higgs and of many possible phenomena beyond the Standard Model.

**Ackowledgments**

The author wishes to thank J.Boyd, A.Di Ciaccio, L.Di Ciaccio, D.Costanzo, R.Goncalo, F.Gianotti, K.Jakobs, P.Jenni, T.Kawamoto, I.Korolkov, W.Kozanecki, T.Le Compte, M.Nessi, L.Nisati, M.Smizanska and other colleagues.

**References**

1 The ATLAS Collaboration, G. Aad et al., "The ATLAS Experiment at the CERN Large Hadron Collider", JINST (2008) S08003.
2 The ATLAS Collaboration, "ATLAS HLT DAQ and Controls", CERN/LHCC/2003-022 (2003).
3 The ATLAS Collaboration, D.Dannheim, "Commissioning and Performance of the ATLAS Liquid Argon Calorimeter", these proceedings.
4 The ATLAS Collaboration, F.Martin, "Commissioning of the ATLAS Inner Tracking Detector", these proceedings.
5 The ATLAS Collaboration, A.Belloni, "Integration and Commissioning of the ATLAS Muon Spectrometer", these proceedings.
6 M.Abolins et al, "Integration of the trigger and data acquisition systems in ATLAS", J.Phys.Conf.Ser.119:022014, 2008.
7 The ATLAS Collaboration, T.Bold, "Commissioning of the ATLAS Trigger", these proceedings.
8 The ATLAS Collaboration, "Expected Performance of the ATLAS Experiment, Detector, Trigger and Physics", CERN-OPEN-2008-020, Geneva, 2008, to appear.